\documentclass[lettersize,journal]{IEEEtran}
\usepackage{amsmath,amsfonts}
\usepackage{algorithmic}
\usepackage{algorithm}
\usepackage{array}
\usepackage[caption=false,font=normalsize,labelfont=sf,textfont=sf]{subfig}
\usepackage{textcomp}
\usepackage{stfloats}
\usepackage{url}
\usepackage{indentfirst}
\usepackage{amsthm}
\theoremstyle{definition}

\usepackage{verbatim}
\usepackage[colorlinks,urlcolor=blue,linkcolor=blue,citecolor=blue]{hyperref}
\newtheorem{theorem}{Theorem}

\usepackage{graphicx}
\usepackage{booktabs}
\usepackage{cite}
\usepackage{soul}
\begin{document}

\title{Leveraging Order-Theoretic Tournament Graphs for Assessing Internal Consistency in Survey-Based Instruments Across Diverse Scenarios}

\author{Muhammad Umair Danish,~\IEEEmembership{Student Member,~IEEE}, 
        Umair Rehman,~\IEEEmembership{Member,~IEEE}, 
        and Katarina Grolinger,~\IEEEmembership{Member,~IEEE}%
\thanks{Muhammad Umair Danish and Katarina Grolinger are with the Department of Electrical and Computer Engineering, The University of Western Ontario, London, ON, Canada (e-mails: \{mdanish3, kgroling\}@uwo.ca).}%
\thanks{Umair Rehman is with the Department of Computer Science, The University of Western Ontario, London, ON, Canada (e-mail: urehman6@uwo.ca).}%
\thanks{Corresponding author: Katarina Grolinger (e-mail: kgroling@uwo.ca).}%
\thanks{This work was supported by the Canada Research Chairs Program under Grant CRC-2022-00078 (K. Grolinger), NSERC Discovery Grant RGPIN-2018-06222 (K. Grolinger), SSHRC Insight Development Grant File No. 430-2024-01140 (U. Rehman), and NSERC Discovery Grant RGPIN-2024-05191 (U. Rehman). Computation was enabled in part by the Digital Research Alliance of Canada.}}%


\maketitle
\markboth{Under Review}{Under Review}

\begin{abstract}

This paper introduces Monotone Delta ($\delta$), an order-theoretic measure designed to enhance the reliability assessment of survey-based instruments in human-machine interactions. Traditional reliability measures, such as Cronbach’s Alpha and McDonald’s Omega, often yield misleading estimates due to their sensitivity to redundancy, multidimensional constructs, and assumptions of normality and uncorrelated errors. These limitations can compromise decision-making in human-centric evaluations, where survey instruments inform adaptive interfaces, cognitive workload assessments, and human-AI trust models. Monotone Delta addresses these issues by quantifying internal consistency through the minimization of ordinal contradictions and alignment with a unidimensional latent order using weighted tournaments. Unlike traditional approaches, it operates without parametric or model-based assumptions. We conducted theoretical analyses and experimental evaluations on four challenging scenarios: tau-equivalence, redundancy, multidimensionality, and non-normal distributions, and proved that Monotone Delta provides more stable reliability assessments compared to existing methods. The Monotone Delta is a valuable alternative for evaluating questionnaire-based assessments in psychology, human factors, healthcare, and interactive system design, enabling organizations to optimize survey instruments, reduce costly redundancies, and enhance confidence in human-system interactions.
\end{abstract}



\section{Introduction}

\IEEEPARstart{R}{eliability} assessment is essential for evaluating survey-based instruments used in human-centered domains such as human-robot collaboration \cite{hoffman2019evaluating}, healthcare \cite{lu2024wearable}, AI-generated content evaluation \cite{10869828} and education \cite{martin2021future}. The assessment ensures that elements within survey questionnaires are internally consistent and collectively measure the intended construct, affirming the truthfulness of the collected data \cite{hayes2020use, ahmed2021reliability}. Internal consistency is a type of reliability assessment that measures how well items in a survey-based instrument contribute to evaluating the same latent construct. Thus, internal consistency is essential for assessing any measuring methodology for collecting primary data. It also enhances the validity of research findings and strengthens their credibility, making it essential for rigorous scientific inquiry \cite{tavakol2011making, karakaya2022sample}. 


Internal consistency assessment is fundamental in human-centered systems to ensure the reliability and accuracy of survey-based instruments used in human-machine interaction, cognitive systems, and AI-assisted decision support \cite{10869828, mu2011networking}. For instance, survey-based evaluations that inform strategies in user experience design, trust in automation, and AI-driven decision systems must reliably capture the constructs they intend to measure. This ensures that each item in a questionnaire meaningfully contributes to its intended construct, enhancing internal consistency and optimizing the instrument's dimensional structure \cite{forero2024cronbach}. Moreover, robust reliability measures mitigate the risks of flawed conclusions by addressing issues such as redundant items, correlated errors, or multidimensional constructs, which can obscure accurate reliability scores and lead to biased managerial decisions \cite{stadler2021knowledge}. For practitioners, this translates into improved confidence in research findings and their applicability to real-world challenges \cite{moenaert1995r}.


Existing reliability measures such as Cronbach's Alpha and McDonald's Omega are widely used to assess the internal consistency of survey-based instruments \cite{barbera2020clarity}. However, these methods are built on assumptions that usually fail in practical applications. For instance, Cronbach's Alpha assumes tau-equivalence, which requires all items to represent the latent construct equally, an assumption rarely met in real-world datasets \cite{barbera2020clarity}. It is also sensitive to redundancy, artificially inflating reliability when similar items are included \cite{agbo2010cronbach, 10869828}. McDonald's Omega addresses some limitations by allowing for unequal item contributions, but still, it relies heavily on factor models sensitive to small sample sizes and uneven data distributions \cite{hayes2020use}. Both measures also depend on assumptions of normality and uncorrelated errors, which are frequently violated in multidimensional or heterogeneous datasets \cite{stensen2022internal}. Addressing these gaps requires a fundamentally different technique that avoids reliance on parametric assumptions, adapts to diverse data distributions, and ensures stability across varying nature of data. The need for an assumption-free, scalable, and robust reliability measure presents an opportunity to advance internal consistency assessment and improve reliability evaluation across disciplines.



 To address the challenges of traditional reliability measures, this paper proposes Monotone Delta $\delta$, an order-theoretic method designed to assess internal consistency by leveraging ordinal relationships among item responses. The core principle of Monotone Delta is to minimize ordinal contradictions in data by arranging responses along an optimal unidimensional latent order. This involves constructing a weighted tournament graph that captures pairwise dominance relationships between items and respondents. Monotone Delta identifies and resolves contradictions to quantify the alignment of responses with a coherent latent structure using the tournament graph technique. The proposed Monotone Delta operates without relying on assumptions such as tau-equivalence, normality, or factor models. Our technique is fundamentally and theocratically different from Cronbach's Alpha and McDonald's Omega because it is based on order theory \cite{davey2002introduction, chen2020prescribed}, which makes it resilient against redundant items, multidimensional constructs, and distributional irregularities. This paper evaluates Monotone Delta through a human-centered study on AI-generated image assessments. The proposed method remains adaptable to other domains utilizing survey-based instruments. The main contributions of this paper are as follows:

\begin{enumerate} 

\item Design of Monotone Delta, an order-theoretic measure quantifying internal consistency by minimizing ordinal contradictions, operating without parametric assumptions, and ensuring robustness against multidimensionality, redundancy, and data irregularities.

\item Design of a systematic evaluation to assess reliability measures under challenging conditions, including tau-equivalence, redundancy, multidimensionality, and non-normal distributions.

\item Theoretical evaluation of Monotone Delta, Cronbach's Alpha, and McDonald Omega, proving Monotone Delta's resilience to multidimensionality, redundancy inflation, and non-normality.

\item Experimental comparison of Monotone Delta with traditional measures, verifying its stable performance across diverse scenarios.
\end{enumerate}

The remainder of the paper is organized as follows: Section \ref{sec:relwork} presents the formal constructs and discusses limitations of existing measures, Section \ref{sec:MoncDelta} details the proposed Monotone Delta, Section \ref{sec:experiments} presents the evaluation, and Section \ref{sec:conclusion} concludes the paper.

\begin{table*}[!t]
\centering
\caption{Comparison of Reliability Measures}
\label{tab:reliability_comparison}
\begin{tabular}{@{}p{4cm} p{3cm} p{3cm} p{3cm}@{}}
\toprule
\textbf{Criterion} & \textbf{Cronbach’s $\alpha$} & \textbf{McDonald’s $\omega$} & \textbf{Monotone $\delta$ (Proposed)}\\ 
\midrule
\textbf{Assumptions}          & Tau-equivalence required & Factor model assumptions & None \\ 
\textbf{Handling Multidimensionality}  & Produces misleading results & Moderately sensitive & Robust against violations \\ 
\textbf{Sensitivity to Item Redundancy}      & Inflates reliability scores        & Overestimates reliability & Resilient to redundancy \\ 
\textbf{Model Dependence}     & No explicit model required            & Relies on factor models          & Independent of parametric models \\ 
\textbf{Robustness to Non-Normality}        & Limited robustness         & Susceptible to deviations   & Fully robust \\ 
\textbf{Computational Complexity}           & Low             & Moderate      & Moderate  \\ 
\bottomrule
\end{tabular}
\end{table*}

\section{Formal Constructs and  Limitations of Existing Reliability Measures} \label{sec:relwork}
This section describes data and variable representation and provides theoretical evidence of challenges associated with Cronbach's Alpha and McDonald's Omega. This section also describes existing alternative methods and introduces Order Theory as a foundation for our work. 

\subsection{Data and Variable Representation} \label{ref:dataVR}
Let $R = \{r_1, r_2, \dots, r_N\}$ represent the set of $N$ respondents, where $r_j$ denotes a single respondent, and let $I = \{i_1, i_2, \dots, i_K\}$ be the set of $K$ items or questions in the survey-based instrument. Each respondent $r_j$ provides a vector of responses:
\begin{equation}
\mathbf{x}_j = (x_{j1}, x_{j2}, \ldots, x_{jK}) \in \mathbb{R}^K,
\end{equation}
where $x_{j\ell}$ denotes the response of respondent $r_j$ to item $i_\ell$. We define the response vector across all respondents for a fixed item denoted as \( \mathbf{X}_\ell\),  representing the set of all responses to an item $i_\ell$ from the $N$ respondents, as:

\begin{equation}
\mathbf{X}_\ell = (x_{1\ell}, x_{2\ell}, \dots, x_{N\ell}) 
\label{eq:2}
\end{equation}
\noindent The total response score for each item $i_\ell$, aggregating responses from all respondents, is given by:
\begin{equation}
X_\ell = \sum_{j=1}^N x_{j\ell}.
\end{equation}

\noindent Here, $x_{j\ell}$ represents individual responses, $\mathbf{X}_\ell$ denotes the vector of responses for the item $i_\ell$ across all respondents, and $X_\ell$ is the aggregate score for the item $i_\ell$. 

\subsection{Cronbach's Alpha}
Cronbach's Alpha, denoted by $\alpha$, is the most widely used method for measuring the internal consistency of a survey-based instrument \cite{kennedy2022sample}. It is defined as:
\begin{equation}
\alpha = \frac{k}{k - 1} \left( 1 - \frac{\sum_{\ell=1}^k \sigma_{X_\ell}^2}{\sigma_T^2} \right),
\end{equation}
where $k$ represents the number of items, $\sigma_{X_\ell}^2$ is the variance of responses for item $i_\ell$, and $\sigma_T^2$ is the variance of the total composite score $T$, which is defined as:
\begin{equation}
T = \sum_{\ell=1}^k X_\ell.
\end{equation}

Cronbach's Alpha is an excellent measure, but it has several limitations, including its assumption of tau-equivalence and the requirement that all items have equal true-score variances. This assumption implies:
\begin{equation}
\text{Cov}(X_\ell, X_m) = \text{Var}(X_\ell), \quad \forall \ell, m \in \{1, 2, \dots, k\},
\end{equation}
where $\text{Cov}(X_\ell, X_m)$ is the covariance between items $i_\ell$ and $i_m$, and $\text{Var}(X_\ell)$ is the variance of item $i_\ell$. However, tau-equivalence rarely holds in practice, as items may differ in their measurement properties, leading to biased estimates. 

The second issue with Cronbach's Alpha is that it is sensitive to the number of items. As the number of items \( k \) increases, the value of \( \alpha \) approaches one, even if the additional items are redundant or do not enhance the quality of the instrument. This behavior can be described as follows:
\begin{equation}
\alpha \to 1 \quad \text{as} \quad k \to \infty.
\end{equation}

When the survey-based instrument captures multiple latent constructs \cite{forero2024cronbach}, the covariance matrix $\Sigma$ of the item responses becomes block-diagonal:
\begin{equation}
\Sigma = \begin{bmatrix}
\Sigma_1 & 0 \\
0 & \Sigma_2
\end{bmatrix},
\end{equation}
where $\Sigma_1$ and $\Sigma_2$ represent covariances within subsets of items measuring distinct constructs. This violates the assumption of unidimensionality, resulting in misleading reliability estimates. These limitations show that while Cronbach's Alpha is widely used, its assumptions and sensitivity to specific conditions restrict its effectiveness as a universal reliability measure.  
\subsection{McDonald Omega}

McDonald's Omega, denoted as \( \omega \), is the second most used technique after Chronback Alpha \cite{hayes2020use, orccan2023comparison, cho2021neither}: it quantifies internal consistency by partitioning the total score variance into variance explained by a common latent factor and unique item variances. The total score \( T_j \) for respondent \( r_j \) is:
\begin{equation}
T_j = \sum_{\ell=1}^K x_{j\ell},
\end{equation}
where \( x_{j\ell} \) represents the response of respondent \( r_j \) to item \( i_\ell \). The variance of the total score \( T_j \) is expressed as:
\begin{equation}
\sigma_T^2 = \sum_{\ell=1}^K \lambda_\ell^2 \sigma_F^2 + \sum_{\ell=1}^K \sigma_{\epsilon_\ell}^2,
\end{equation}
where \( \lambda_\ell \) denotes the factor loading of item \( i_\ell \), \( \sigma_F^2 \) represents the variance of the common latent factor \( F_j \), and \( \sigma_{\epsilon_\ell}^2 \) denotes the unique variance of item \( i_\ell \). McDonald's Omega is formally defined as:
\begin{equation}
\omega = \frac{\sum_{\ell=1}^K \lambda_\ell^2 \sigma_F^2}{\sigma_T^2}.
\end{equation}
This expression measures the proportion of total variance in the responses attributable to the common latent factor \( F_j \). The common latent factor \( F_j \) represents the shared variance across all measurement instrument items, reflecting the measured construct. The unique variances (\( \sigma_{\epsilon_\ell}^2 \)) correspond to item-specific variability not explained by the common factor, and these are assumed to be uncorrelated across items:
\begin{equation}
\text{Cov}(\epsilon_{j\ell}, \epsilon_{jm}) = 0 \quad \text{for} \quad \ell \neq m.
\end{equation}

The limitations of McDonald's Omega arise from specific factor model assumptions inherent in its computation, and Uncorrelated errors are often violated in practice. The overlapping content among items refers to items that assess highly similar aspects of a construct and can introduce error correlations, which leads to biased estimates of \( \omega \):
\begin{equation}
\text{Cov}(\epsilon_{j\ell}, \epsilon_{jm}) \neq 0 \quad \text{for} \quad \ell \neq m.
\end{equation}

Weak factor loadings \( (\lambda_\ell \approx 0) \) reduce the contribution of items to the numerator:
\begin{equation}
\sum_{\ell=1}^K \lambda_\ell^2 \sigma_F^2,
\end{equation}
This disproportionately inflates the denominator due to increased unique variance, leading to underestimated reliability.

Redundancy among items inflates the total score variance \( \sigma_T^2 \) for items measuring identical constructs. The variances compound, resulting in poor reliability. Such inflation artificially raises \( \omega \), undermining its interpretive value.

Moreover, in multidimensional datasets, items may correspond to distinct latent factors, leading to a block-diagonal covariance structure akin to the \( \omega \)-specific form:

\begin{equation}
\Sigma = \begin{bmatrix}
\Sigma_{\phi_1}\Theta_1 & 0 \\
0 & \Sigma_{\phi_2}\Theta_2
\end{bmatrix},
\end{equation}

where \( \Sigma_{\phi_1} \) and \( \Sigma_{\phi_2} \) represent the factor covariance matrices for two latent dimensions, and \( \Theta_1 \) and \( \Theta_2 \) denote their respective residual variances. This structure violates the unidimensionality assumption, potentially rendering \( \omega \) an inadequate measure of reliability.

From the discussed challenges, which are also summarized in Table \ref{tab:reliability_comparison}, it is evident that both measures have limitations, which limit their applicability across a wide range of applications. As the sophistication of questionnaire development continues to evolve, there is an urgent need for new measures to address the challenges both techniques face.

\subsection{Alternative Methods}
In addition to Cronbach's Alpha and McDonald's Omega, other techniques such as Greatest Lower Bound (GLB) and Split-Half Reliability have been proposed as an alternative measure of internal consistency. The GLB \cite{ten2004greatest, cho2022reliability} estimates reliability by optimizing the covariance matrix of items. The GLB usually outperforms Cronbach's Alpha, but it requires intensive computational resources and fails to address redundancy and multidimensional data. Moreover, its reliance on matrix optimization limits scalability to large datasets \cite{ten2004greatest, cho2022reliability}. Split-Half Reliability \cite{chakrabartty2013best} is a notable measure that partitions items into two subsets and evaluates the correlation between their scores. Despite its simplicity, the method is sensitive to how items are divided, leading to variability in reliability estimates. This measure also does not consider ordinal relationships, which is a considerable limitation in datasets with ties or noise.  

While alternative measures such as GLB and Split-Half Reliability present more options for assessing internal consistency, they share common limitations due to their theoretical reliance on Cronbach's Alpha and McDonald's Omega. These techniques extend or modify either Cronbach's Alpha and McDonald's Omega; for example, GLB refines Cronbach's Alpha through covariance matrix optimization, and Split-Half Reliability evaluates subset correlations and simplifies Omega by focusing on inter-item relationships. However, their shared assumptions, including unidimensionality and pairwise independence of items, limit their applicability to handle redundancy, noise, and multidimensionality. Given the widespread usage and theoretical prominence of Cronbach's Alpha and McDonald's Omega, they remain the most impactful benchmarks for comparison. We address these gaps by introducing a novel order-theoretic method that explicitly quantifies contradictions and incorporates robust handling of ties and noise, delivering a more reliable and scalable solution for modern datasets.

\subsection{Order Theory} 
We employ order theory as a foundation to overcome the limitations of traditional reliability measures. It provides a mathematical framework for analyzing hierarchical and sequential relationships, such as greater than, less than, and precedes \cite{davey2002introduction, chen2020prescribed}. By formalizing these intuitive relationships through the lens of partial orders, this framework provides a robust mechanism for evaluating ordering and coherence within datasets. A partial order constitutes a binary relation \( \preceq \) on a set \( P \) that adheres to three fundamental properties: 
\begin{align} a \preceq a & \quad \text{(reflexivity)}, \\ a \preceq b \text{ and } b \preceq a & \implies a = b \quad \text{(antisymmetry)}, \\ a \preceq b \text{ and } b \preceq c & \implies a \preceq c \quad \text{(transitivity)}. \end{align} 
In the field of measurement instruments, the response set \( R \) and items \( I \) establish a partially ordered set (poset) when their responses reflect an inherent order based on a latent trait. For example, higher scores typically signify a greater alignment with the measured construct. Order-preserving (monotone) functions are pivotal in evaluating the internal consistency of measurement instruments. A function \( f : P \to Q \) is considered monotone if it satisfies the condition: 

\begin{equation} a \preceq b \implies f(a) \preceq f(b). \end{equation} 
Monotonicity ensures the preservation of the latent ordering of responses under transformations, thereby facilitating meaningful interpretations of aggregated scores. Contradictions arise when observed responses deviate from the assumed latent ordering. For a poset \( P \) with relation \( \preceq \), these contradictions become evident through pairs \( (a, b) \in P \times P \) such that: \begin{equation} a \preceq b \quad \text{and} \quad b \prec a. \end{equation} These violations disrupt the dataset's unidimensionality, complicating the interpretation of reliability measures. Addressing these contradictions is critical for deriving reliable internal consistency estimates. 

To solve challenges faced by both Chronback Alpha and McDoland Omega, we aim to employ the principles of order theory to quantify internal consistency by minimizing ordinal contradictions. The order theory can assess the alignment of item responses with a latent order, defined by a poset \( P \) in which items \( i_\ell \) and responses \( x_{j\ell} \) fulfill the requirement: 

\begin{equation} x_{j\ell} \preceq x_{jm} \implies i_\ell \preceq i_m. \end{equation}

This ordinal relationship enables practical evaluation of internal consistency across complex and heterogeneous questionnaires.

\section{Monotone Delta}\label{sec:MoncDelta}
This section describes the proposed Monotone Delta, including the Theoretic Formulation of Monotone Delta, Monotone Delta Definition and Normalization, its properties, and theoretical examination.

\subsection{Theoretic Formulation of Monotone Delta}
We use the notation for \( R \), \( I \), and \( \mathbf{x}_j \) as defined in Subsection \ref{ref:dataVR} and introduce additional symbols. Let \( \pi \) denote a permutation that orders respondents based on their responses. The function \( W(j, k) \) counts the number of items where respondent \( r_j \) outperforms \( r_k \), and it is used to construct weighted tournaments. A weighted tournament refers to a type of directed graph used in order theory to represent pairwise relationships among elements, such as respondents or items \cite{connelly2014tournament,rajkumar2021theory}.   The symbol \( C(\pi) \) represents the total contradiction count for a given ordering \( \pi \), quantifying deviations from the latent order. 

The purpose is to evaluate how well the responses align with a single monotone latent dimension by minimizing contradictions. Consider a poset \( (R, \preceq) \), where \( \preceq \) represents a hypothesized latent order that reflects the unidimensional trait being measured. The goal is to align the respondents \( R = \{r_1, r_2, \dots, r_N\} \) with this latent order. Let \( \pi: \{1, \dots, N\} \to \{1, \dots, N\} \) denote a permutation that provides a linear extension of the poset, meaning the respondents are arranged such that:
\begin{equation}
r_{\pi(1)} \preceq r_{\pi(2)} \preceq \cdots \preceq r_{\pi(N)}.
\end{equation}

The respondents' responses should respect this ordering if the data are perfectly unidimensional and free of noise. For any pair of respondents \( j \) and \( k \) where \( \pi(j) < \pi(k) \), the responses for all items should satisfy:
\begin{align}\label{eq:ideal_order}
\pi(j) < \pi(k) &\implies x_{\pi(j)\ell} \le x_{\pi(k)\ell}, \notag \\
&\quad \forall \ell \in \{1, \dots, K\}.
\end{align}

\noindent Here  \( x_{\pi(j)\ell} \) represents the response of the \( j \)-th respondent (according to the permutation \( \pi \)) to the \( \ell \)-th item. The inequality \( x_{\pi(j)\ell} \le x_{\pi(k)\ell} \) implies that respondent \( r_{\pi(j)} \) shows a response no stronger than respondent \( r_{\pi(k)} \) for all items, consistent with the hypothesized latent order. Contradictions occur due to multidimensionality, noise, or redundant patterns. A contradiction is defined as a violation of Equation (\ref{eq:ideal_order}), i.e., there exists $j<k$ and an item $\ell$ such that:
\begin{equation}\label{eq:contradiction_condition}
x_{\pi(j)\ell} > x_{\pi(k)\ell}.
\end{equation}

The degree of contradiction measures how far the data deviates from a perfect unidimensional ordering. To quantify contradictions in respondent scores, we use the concept of a "weighted tournament," a directed graph where vertices correspond to respondents, and directed edges indicate dominant relationships based on their responses. The edge weights quantify in how many items one respondent outperforms another, and this computes the analysis of pairwise contradictions and the optimization of respondent orderings. The function \( W(j, k) \) is defined as:
\begin{equation}\label{eq:W_function}
W(j, k) = \#\{\ell : x_{j\ell} > x_{k\ell}\},
\end{equation}
where \( W(j, k) \) represents the number of items (\( \ell \)) for which respondent \( r_j \) scores higher than respondent \( r_k \).  The symbol \( \# \{\dots\} \) denotes the cardinality of the set (i.e., the count of elements in the set). For example, if respondent \( r_j \) scores higher than \( r_k \) on 3 out of 5 items, then \( W(j, k) = 3 \). This structure induces a \emph{weighted tournament} on \( N \) vertices, with directed edges weighted by \( W(j, k) \).

To analyze contradictions, we consider a linear extension of the poset, a specific ordering \( \pi \) of respondents that respects the poset's partial order as much as possible. A linear extension arranges respondents \( r_1, \dots, r_N \) in a total order, such that if \( r_j \preceq r_k \) in the poset, then \( r_j \) appears before \( r_k \) in \( \pi \). However, due to noise or multidimensionality, the responses may not perfectly align with the poset's partial order, resulting in contradictions. For a given ordering \( \pi \), the total contradiction count is:
\begin{equation}\label{eq:C_pi}
C(\pi) = \sum_{1 \leq j < k \leq N} \#\{\ell : x_{\pi(j)\ell} > x_{\pi(k)\ell}\}.
\end{equation}
This equation counts the number of item-level violations of the ordering \( \pi \). A contradiction occurs when \( x_{\pi(j)\ell} > x_{\pi(k)\ell} \) despite \( \pi(j) < \pi(k) \), indicating that respondent \( r_{\pi(j)} \) unexpectedly outperforms \( r_{\pi(k)} \) on some items. To find the optimal ordering, we iteratively refine \(\pi\) by evaluating pairwise swaps of respondents and accepting swaps that reduce the contradiction count \( C(\pi) \). The process continues until \( C(\pi) \) converges to its minimum value \( C^* \). 

\begin{equation}\label{eq:C_star}
C^* = \min_{\pi} C(\pi).
\end{equation}
The optimal ordering \( \pi^* \), obtained through this refinement, aligns responses as closely as possible to the hypothesized latent order. This method is equivalent to solving a minimum feedback arc set problem \cite{younger1963minimum} on the weighted tournament defined by \( W(j, k) \).

\subsection{Monotone Delta Definition and Normalization}
The maximum possible contradiction count, $C_{\max}$, occurs if for every pair $(r_j, r_k)$ with $j<k$, the ordering $\pi$ is reversed relative to their observed dominance. Each pair can contribute up to $K$ contradictions, and there are $N(N-1)/2$ pairs, thus:
\begin{equation}\label{eq:C_max}
C_{\max} = K \cdot \frac{N(N-1)}{2}.
\end{equation}

We define the Monotone Delta as:
\begin{equation}\label{eq:MOR_def}
\delta = 1 - \frac{C^*}{C_{\max}}.
\end{equation}

The value \(\delta = 1\) indicates perfect unidimensional coherence, while lower values of \(\delta\) reflect weaker coherence due to increased contradictions. As the dataset complexity increases (e.g., multiple latent dimensions, correlated errors, redundant items), $C^*$ increases, reducing $\delta$ and signaling weaker unidimensional coherence. The ties in \( W(j, k) \) and noise (\( \epsilon_{j\ell} \)) are handled by ensuring they do not artificially inflate \( C^* \) by maintaining the reliability. Algorithm:  \ref{alg:monotone_delta} describes all the computational steps of the proposed Monotone Delta ($\delta$).

\subsection{Properties and Theoretical Results}

\begin{theorem}[Scale Invariance]\label{thm:scale_invariance}
Consider any strictly increasing transformation 
\(\; g_{\ell}: \mathbb{R} \rightarrow \mathbb{R}\;\) 
applied item wise, i.e., 
\(\; x_{j\ell} \mapsto g_{\ell}(x_{j\ell}).\)
Then, the relative ordering among responses is preserved, implying that:
\begin{equation}
C(\pi), \quad C^*, \quad \text{and} \quad \delta 
\quad\text{ are unaffected by}\quad g_{\ell}.
\end{equation}
\end{theorem}

\begin{proof}
Since \(g_{\ell}\) is strictly increasing, we have 
\begin{equation}
x_{j\ell} > x_{k\ell} 
\;\Longleftrightarrow\; 
g_{\ell}\bigl(x_{j\ell}\bigr) > g_{\ell}\bigl(x_{k\ell}\bigr).
\end{equation}

\noindent No new contradictions can be introduced or removed by such transformation. The structural properties of the weighted tournament (and thus the minimal contradiction count) remain unchanged. Therefore, \(\delta\) is unaffected by scale changes. Further discussion on scale invariance in ordinal methods can be found in~\cite{bowen2015conducting}.
\end{proof}

\noindent \textit{This proof verifies that Monotone Delta remains robust to scale changes, unlike Cronbach's Alpha, which is sensitive to such transformations \cite{agbo2010cronbach}.}

\begin{theorem}[Sensitivity to Multidimensionality]\label{lemma:multidim_sensitivity}
Let there be \(d>1\) latent dimensions, each affecting a distinct subset of items 
\(\;I = I_{1}\cup I_{2}\cup \dots \cup I_{d}\).
If these dimensions are sufficiently distinct, then for large \(N\) there exists 
\(\beta(N,K,d) > 0\) such that
\begin{equation}\label{eq:multidim_lower_bound}
\mathbb{E}\bigl[C^*\bigr] \;\;\ge\;\; \beta(N, K, d),
\end{equation}
and therefore,
\begin{equation}\label{eq:multidim_delta_ineq}
\delta \;\;\le\;\; 1 \;-\; \frac{\beta(N,K,d)}{C_{\max}}.
\end{equation}
\end{theorem}

\begin{proof}[Proof]
If items are truly governed by multiple dimensions, a single total ordering cannot perfectly satisfy all item-response relations. The resulting “dimension conflicts” impose a positive lower bound on the minimal contradiction count. Formally, one can decompose the weighted tournament into sub-tournaments driven by each dimension and show via the minimum feedback arc set approach that these independent structures force additional contradictions. This returns \(\beta(N,K,d)\) as a lower bound on \(\mathbb{E}[C^*]\).
\end{proof}

\noindent \textit{This indicates that as multidimensional conflicts intensify, Monotone Delta decreases, detecting deviations from unidimensionality that Cronbach's Alpha or McDonald Omega fail to reveal.}

\begin{theorem}[Redundancy Resistance]
If \( r \) redundant items identical (up to small perturbations \(\epsilon\)) to an existing item subset are added, the minimal contradiction count remains stable:
\begin{equation}\label{eq:redundancy_invariance}
C^*(N, K + r) \approx C^*(N, K).
\end{equation}
\end{theorem}

\begin{proof}
Let the response vector for respondent \( r_j \) be \( \mathbf{x}_j = (x_{j1}, x_{j2}, \ldots, x_{jK}) \). Redundant items are defined as:
\begin{align}
x_{j(K+m)} &= x_{j\ell} + \epsilon_{j(K+m)}, \notag \\
m &= 1, \ldots, r, \notag \\
\ell &\in \{1, \ldots, K\}.
\end{align}
where \( \epsilon_{j(K+m)} \) represents small independent perturbations. The updated weight function \( W'(j, k) \) is:
\begin{align}
W'(j, k) &= W(j, k) \notag \\
&\quad + \sum_{m=1}^r \mathbb{I}\bigl(x_{j(K+m)} > x_{k(K+m)}\bigr).
\end{align}

\noindent where \( \mathbb{I}(\cdot) \) is the indicator function. For redundant items, assuming small \(\epsilon_{j(K+m)}\), we have:
\begin{align}
\text{If } x_{j\ell} &> x_{k\ell}, \notag \\
\text{then } x_{j(K+m)} &> x_{k(K+m)}, \quad \forall m.
\end{align}

Thus, redundant items preserve the relative ordering between \( r_j \) and \( r_k \), contributing no additional contradictions. The total contradiction count for an ordering \( \pi \) after adding redundant items is:
\begin{align}
C'(N, K + r) &= C(N, K) \notag \\
&\quad + \sum_{j < k} \sum_{m=1}^r 
\mathbb{I}\bigl(x_{\pi(j)(K+m)} > x_{\pi(k)(K+m)}\bigr).
\end{align}

\noindent Since 
\begin{align}
\mathbb{I}(x_{\pi(j)(K+m)} &> x_{\pi(k)(K+m)}) \notag \\
&= \mathbb{I}(x_{\pi(j)\ell} > x_{\pi(k)\ell}).
\end{align}

\noindent the contradictions remain unchanged:
\begin{equation}
C'(N, K + r) = C(N, K).
\end{equation}

\noindent Thus, the minimal contradiction count satisfies:
\begin{equation}
C^*(N, K + r) = C^*(N, K).
\end{equation}
For small \(\epsilon\), the perturbations introduced by redundant items are negligible, ensuring:
\begin{equation}
C^*(N, K + r) \approx C^*(N, K).
\end{equation}
\end{proof}
\noindent \textit{This proof verifies that Monotone Delta \(\delta\) is inherently resilient to redundancy, unlike traditional measures such as Alpha or Omega, which inflate reliability scores when redundant items are added.}

\begin{algorithm}[!t]
\caption{Monotone Delta ($\delta$)}\label{alg:monotone_delta}
\begin{algorithmic}[1]
\STATE \textbf{Input:} Response matrix $X \in \mathbb{R}^{N \times K}$, where $x_{j\ell}$ is the response of respondent $r_j$ to item $i_\ell$.
\STATE \textbf{Output:} Monotone Delta $\delta$, a measure of internal consistency in $[0,1]$.

\vspace{0.1cm}

\STATE \textbf{Step 1: Construct Weighted Tournament}
\STATE Initialize a directed, weighted graph $G=(V,E)$ with vertices $V=\{r_1, \ldots, r_N\}$.
\FOR{$j = 1$ to $N$}
    \FOR{$k = 1$ to $N$ with $k \neq j$}
        \STATE Compute $W(j,k) = \#\{\ell \mid x_{j\ell} > x_{k\ell}\}$.
        \STATE Add a directed edge from $r_j$ to $r_k$ with weight $W(j,k)$ to $G$.
    \ENDFOR
\ENDFOR

\vspace{0.1cm}

\STATE \textbf{Step 2: Initial Ordering}
\STATE Compute the mean score for each respondent $r_j$: $\bar{x}_j = \frac{1}{K}\sum_{\ell=1}^K x_{j\ell}$.
\STATE Sort respondents according to $\bar{x}_j$ to obtain an initial permutation $\pi$.

\vspace{0.1cm}

\STATE \textbf{Step 3: Local Search Optimization}
\STATE Set $C(\pi) = \sum_{1 \le j < k \le N} \#\{\ell : x_{\pi(j)\ell} > x_{\pi(k)\ell}\}$.
\STATE \textbf{repeat}
    \STATE \hspace{0.5cm} Select a pair $(r_{\pi(p)}, r_{\pi(q)})$ at random, with $p<q$.
    \STATE \hspace{0.5cm} Create a new permutation $\pi'$ by swapping $r_{\pi(p)}$ and $r_{\pi(q)}$.
    \STATE \hspace{0.5cm} Compute $C(\pi')$. 
    \STATE \hspace{0.5cm} \textbf{if} $C(\pi') < C(\pi)$ \textbf{then} accept $\pi' \leftarrow \pi$ and $C(\pi) \leftarrow C(\pi')$.
\STATE \textbf{until} no improving swap is found after several attempts.

\vspace{0.1cm}

\STATE \textbf{Step 4: Compute Minimal Contradiction Count}
\STATE After convergence, let $\pi^*$ be the final permutation found and $C^* = C(\pi^*)$ be the minimal contradiction count obtained.

\vspace{0.1cm}

\STATE \textbf{Step 5: Calculate Monotone Delta}
\STATE Compute the maximum possible contradiction count $C_{\max} = K \cdot \frac{N(N-1)}{2}$.
\STATE Compute $\delta = 1 - \frac{C^*}{C_{\max}}$.

\vspace{0.1cm}

\STATE \textbf{return} $\delta$

\end{algorithmic}
\end{algorithm}

\begin{table*}[!h]
    \centering
    \caption{Visual Verity Questionnaire}
    \label{tab:photorealism_questionnaire}
    \begin{tabular}{@{}p{3cm}p{11cm}p{2cm}@{}}
    \toprule
    \textbf{Question ID} & \textbf{Question Text} & \textbf{Scale} \\ 
    \midrule
    \multicolumn{3}{@{}l}{\textit{Demographic Questions (DQ)}} \\ 
    DQ1 & What is your gender? & Multiple choice \\ 
    DQ2 & What is your age? & Open-ended \\ 
    DQ3 & What is your educational qualification? & Multiple choice \\ 
    DQ4 & Experience with AI or computer-generated images. & Likert (1–5) \\ 
    DQ5 & Frequency of viewing digital images/graphics. & Likert (1–5) \\ 
    DQ6 & Experience in graphic design or photography. & Yes/No \\ 
    DQ7 & What is your country of residence? & Open-ended \\ 
    \midrule
    \multicolumn{3}{@{}l}{\textit{Photorealism Assessment (PR)}} \\ 
    PR1 & The image looks like a photograph of a real scene. & Likert (1–5) \\ 
    PR2 & I can easily imagine seeing this image in the real world. & Likert (1–5) \\ 
    PR3 & The visual details in this image make it appear realistic. & Likert (1–5) \\ 
    PR4 & The textures in the image look natural and real. & Likert (1–5) \\ 
    PR5 & The lighting and shadows in the image contribute to its realism. & Likert (1–5) \\ 
    \midrule
    \multicolumn{3}{@{}l}{\textit{Image Quality (IQ)}} \\ 
    IQ1 & The image is clear and sharp. & Likert (1–5) \\ 
    IQ2 & The colors in the image are vibrant and lifelike. & Likert (1–5) \\  
    IQ3 & I am satisfied with the overall quality of this image. & Likert (1–5) \\ 
    IQ4 & The image has no visible artifacts or distortions. & Likert (1–5) \\ 
    IQ5 & The resolution of the image meets my expectations. & Likert (1–5) \\ 
    \midrule
    \multicolumn{3}{@{}l}{\textit{Caption Consistency (CC)}} \\ 
    CC1 & The image perfectly aligns with the given caption. & Likert (1–5) \\ 
    CC2 & The elements in the image correspond to the described scene in the caption. & Likert (1–5) \\ 
    CC3 & If I were to describe this image with a caption, it would closely match the provided one. & Likert (1–5) \\ 
    CC4 & The image misses some details mentioned in the caption. & Likert (1–5) \\ 
    CC5 & I feel the image is a true representation of the given caption. & Likert (1–5) \\ 
    \bottomrule
    \end{tabular}
\end{table*}

\section{Experimental Evaluation} \label{sec:experiments}
This section describes a human study and an evaluation through four scenarios and computational complexity. 

\subsection{Human Study}
We designed and conducted a human subject study named Visual Verity, with a sample size of 350 participants for AI-generated images. The AI-generated image dataset was chosen for its direct impact on managerial decisions in engineering management, including marketing, product design, and strategic innovation. The study consists of 22 questions assessing four distinct constructs to evaluate the perceptual quality and experiential responses to AI-generated images from three commercial models and camera-captured images. We got ethics approval from the Non-Medical Research Ethics Board at Western University Ontario to ensure ethical compliance in participant recruitment and data collection. We recruited participants via an online platform, namely prolific \cite{albert2023comparing}, which is known for its diverse variety of pool.

The dataset evaluates images generated by three commercial AI models - DALL·E 3, DALL·E 2, and Stable Diffusion - and camera-captured images. These models represent different strategies for image generation and provide a diverse range of outputs regarding photorealism, coherence, and quality. The questionnaire given in (Table~\ref{tab:photorealism_questionnaire}) assesses multiple dimensions of image evaluation: demographics, photorealism, image quality, and caption consistency. It uses a mix of Likert-scale, multiple-choice, and open-ended questions designed to gather comprehensive feedback from participants. 

\begin{figure}[!h]
    \centering
    \includegraphics[width=\columnwidth]{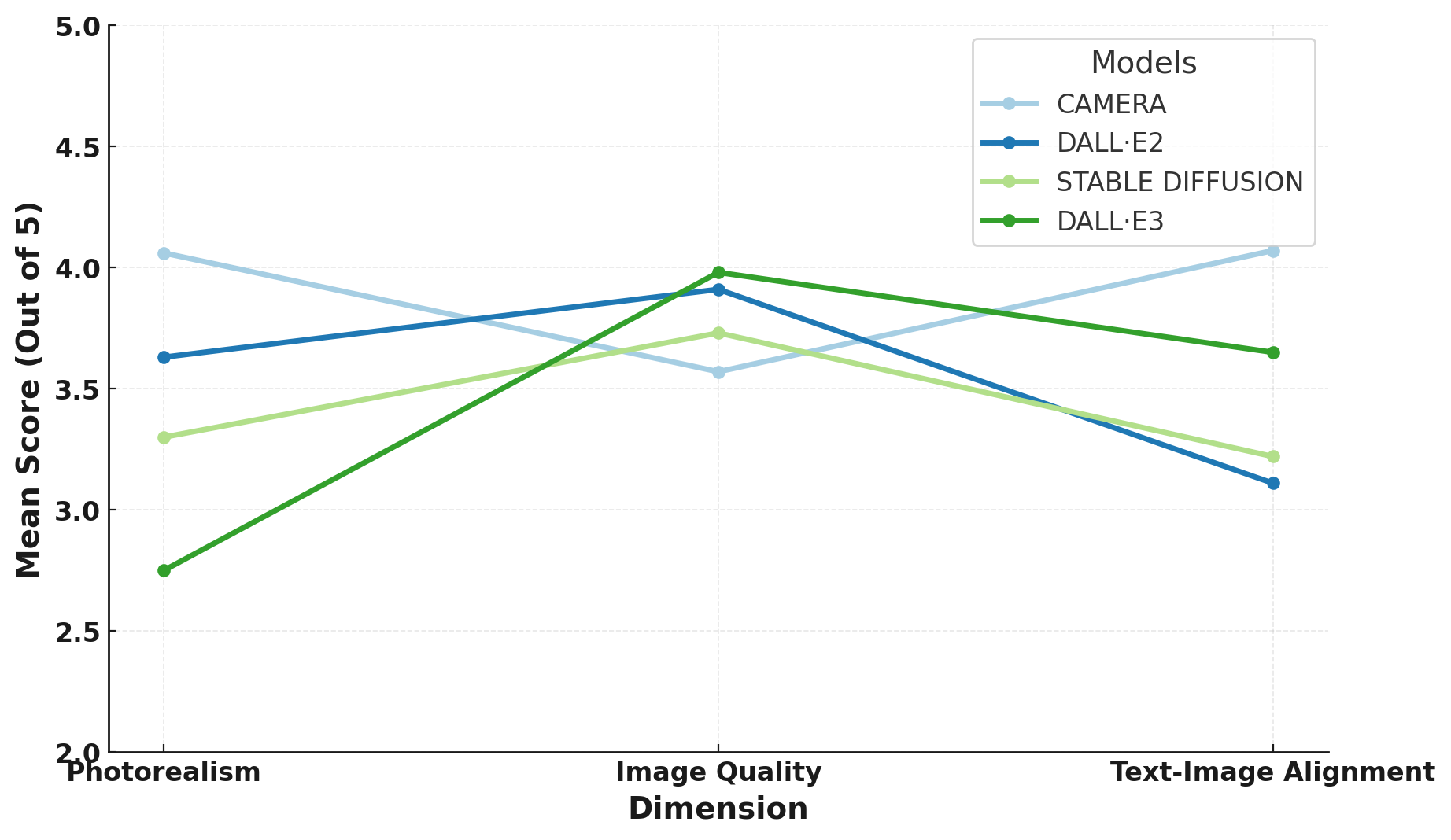}
    \caption{Comarison of participant responses across models (Camera, DALL·E2, Stable Diffusion, and DALL·E3) for three evaluation dimensions: Photorealism, Image Quality, and Text-Image Alignment. DALL·E3 and Stable Diffusion show contrasting trends in Image Quality, while Camera scores consistently high across dimensions.}
    \label{fig:results}
\end{figure}

The questionnaire is a reliable foundation for internal consistency experiments due to its diversity and complexity; for example, data was collected against four constructs, totaling 67 questions and allowing us to assess response alignment and coherence across multiple evaluation dimensions. We have presented results in Figure \ref{fig:results} and Table \ref{tab:overall} shows overall average results that show Camera images are highly realistic, achieving the highest scores in photorealism and text-image alignment. Since the purpose of this paper is to assess the internal consistency of the questionnaire, this assessment will focus less on questionnaire results but will emphasize examining internal consistency.


\begin{table}[!h]
\setlength{\tabcolsep}{3pt}
    \centering
    \caption{Mean Participant Responses (Out of 5)}
    \label{tab:overall}
    \begin{tabular}{lccccc}
    \toprule
        \textbf{Dimension} & \textbf{Camera} & \textbf{DALL·E2} & \textbf{GLIDE} & \textbf{Stable} & \textbf{DALL·E3} \\
         &  &  &  & \textbf{Diffusion} &  \\
        \midrule
        Photorealism & 4.06 & 3.63 & 2.04 & 3.30 & 2.75 \\ 
        Image Quality & 3.57 & 3.91 & 2.10 & 3.73 & 3.98 \\
        Text-Image Align. & 4.07 & 3.11 & 2.03 & 3.22 & 3.65 \\
        \bottomrule
    \end{tabular}
\end{table}

\begin{table*}[!h]
\centering
\caption{Reliability Measures Across Datasets under Ideal Conditions}
\label{tab:reliability_results}
\begin{tabular}{@{}lccccc@{}}
\toprule
\textbf{Dataset} & \textbf{Cronbach's Alpha} & \textbf{McDonald's Omega} & \textbf{GLB} & \textbf{Split-Half Reliability} & \textbf{Monotone Delta} \\ 
\midrule
Camera           & 0.89                      & 0.90                      & 0.79         & 0.71                           & 0.88                     \\
DALL·E2         & 0.94                      & 0.95                      & 0.93         & 0.87                           & 0.92                     \\
DALL·E3         & 0.93                      & 0.94                      & 0.90         & 0.83                           & 0.91                     \\
Stable Diffusion & 0.96                      & 0.97                      & 1.01         & 0.90                           & 0.92                     \\
Overall         & 0.92                      & 0.94                      & 0.78         & 0.86                           & 0.91                     \\
\bottomrule
\end{tabular}
\end{table*}

\subsection{Scenario 1: Tau-Equivalence (Near-Ideal Condition)} 
To establish the validity of our method, Monotone Delta, we first evaluate its performance under ideal conditions and compare it with established baseline measures, Cronbach's Alpha and McDonald's Omega. This will give confidence that the proposed method performs nearly equal to established baseline measures under normal conditions. We also extend comparisons with other measures such as GLB and Split-Half Reliability. Table \ref{tab:reliability_results} summarizes the reliability scores across four datasets such as Camera, DALL·E2, DALL·E3, and Stable Diffusion, and their combined overall dataset. All reliability measures show strong internal consistency, with values close to 1 indicating high reliability and values closer to 0 reflecting weak internal consistency. For the Camera dataset, Cronbach's Alpha scored 0.89, indicating strong internal consistency. McDonald's Omega aligns closely with a score of 0.90, further validating the reliability of the dataset. Monotone Delta, our proposed method, scored 0.88, showing close agreement with Cronbach's Alpha and McDonald's Omega. This alignment with established measures establishes the validity of Monotone Delta under ideal conditions, as it performs similarly to these well-established measures, instilling confidence in its use for further evaluation under more complex scenarios.

\begin{figure}[!t]
    \centering
    \includegraphics[width=\columnwidth]{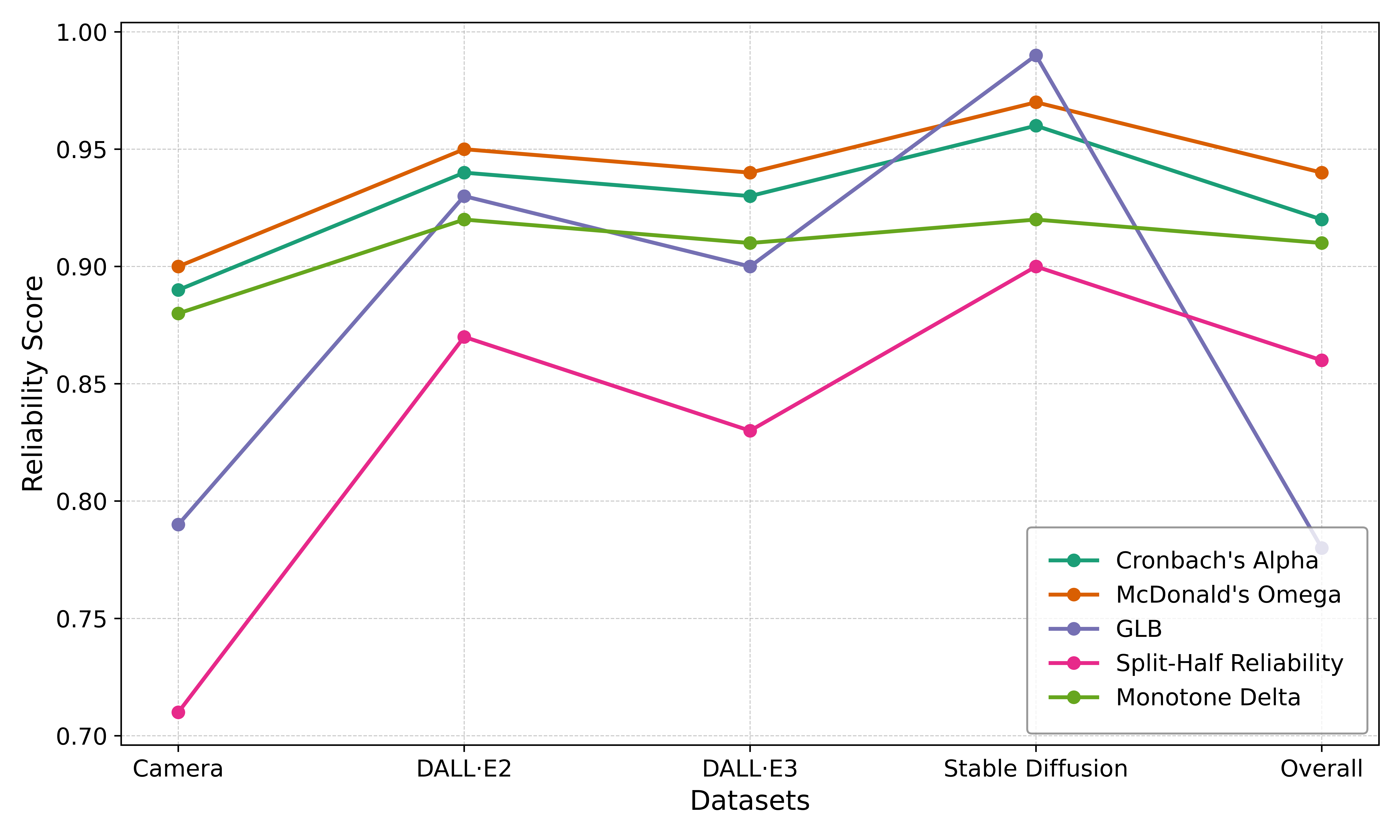} 
    \vspace{-15pt}
    \caption{Readability scores under ideal conditions show that Monotone Delta performs similarly to Chronback's Alpha and McDoland Omega.}
    \label{fig:idealcondition}
\end{figure}

For the DALL·E2 dataset, Cronbach's Alpha reaches a higher value of 0.94, explaining stronger internal consistency. McDonald's Omega closely follows, with a score of 0.95. Monotone Delta also performs similarly in this scenario, achieving a score of 0.92. For the DALL·E3 dataset, Cronbach's Alpha scored 0.93, McDonald's Omega achieved 0.94, and Monotone Delta scored 0.91, reflecting consistent agreement between the three measures. For the stable diffusion and overall dataset, the proposed method performs similarly to the established baselines, which ensures we now perturb our dataset to create another scenario and determine whether Monotone Delta and other measures give stable results or not. 

\begin{table*}[!h]
\centering
\caption{Reliability Measures Across Datasets Under Redundancy Scenario}
\label{tab:scenario2_results}
\begin{tabular}{@{}lccccc@{}}
\toprule
\textbf{Dataset} & \textbf{Cronbach's Alpha} & \textbf{McDonald's Omega} & \textbf{GLB} & \textbf{Split-Half Reliability} & \textbf{Monotone Delta} \\ 
\midrule
Camera           & 0.93                     & 0.94                     & 0.80        & 0.86               & 0.82                 \\
DALL·E2         & 0.96                     & 0.97                     & 0.93        & 0.93               & 0.83                 \\
DALL·E3         & 0.95                     & 0.90                     & 0.88        & 0.86               & 0.85                 \\
Stable Diffusion & 0.98                     & 0.93                     & 1.00        & 0.95               & 0.80                 \\
Overall         & 0.95                     & 0.91                     & 0.79        & 0.83               & 0.84                 \\
\bottomrule
\end{tabular}
\end{table*}
\subsection{Scenario 2: Inflation by Redundant Items}
In this scenario, we manually apply redundancy to the datasets by adding new items that are highly similar to existing ones. The redundant items were generated as linear combinations of original items with a redundancy factor of 0.95, meaning the new items were almost identical to the originals, with a small amount of random noise added. This modification aimed to assess the resilience of reliability measures against inflation caused by redundant items, which artificially increase item correlations and often lead to inflated reliability scores. 

Table \ref{tab:scenario2_results} presents the reliability measures across datasets, such as Cronbach's Alpha, sensitive to the number of items and their correlations, which showed inflated scores across all datasets. For example, in the Stable Diffusion dataset, Cronbach's Alpha increased to 0.98, indicating an artificially high level of internal consistency. This result reflects the measure's susceptibility to redundancy, as adding redundant items leads to overestimating reliability. McDonald's Omega also displayed inflated scores, though to a slightly lesser extent compared to Cronbach's Alpha. In the Stable Diffusion dataset, McDonald's Omega reached 0.93, confirming that it, too, is influenced by redundant items, albeit less dramatically than Cronbach's Alpha. 

GLB also showed inflation under redundancy. For instance, in the Stable Diffusion dataset, GLB scored 1.00, surpassing all other measures and showing strong internal consistency, but it is misleading. Split-half reliability also performed poorly under redundancy, showing varying degrees of inflation. Split-half reliability in the DALL·E2 dataset increased to 0.93, reflecting the influence of redundant items on these measures. Monotone Delta, in contrast, showed resilience to redundancy across all datasets. In the Stable Diffusion dataset, it scored 0.80, closely aligning with its performance under ideal conditions and remaining unaffected by adding redundant items. The score in other datasets' similarity decreases, which shows Monotone Delta's resilience.

\begin{table*}[!t]
\centering
\caption{Reliability Measures Across Datasets Under Multidimensionality Scenario}
\label{tab:scenario3_results}
\begin{tabular}{@{}lccccc@{}}
\toprule
\textbf{Dataset} & \textbf{Cronbach's Alpha} & \textbf{McDonald's Omega} & \textbf{GLB} & \textbf{Split-Half Reliability} & \textbf{Monotone Delta} \\ 
\midrule
Camera           & 0.84                     & 0.86                     & 0.68        & 0.22               & 0.75                 \\
DALL·E2         & 0.85                     & 0.88                     & 0.72        & 0.37               & 0.77                 \\
DALL·E3         & 0.87                     & 0.89                     & 0.75        & 0.41               & 0.78                 \\
Stable Diffusion & 0.89                     & 0.91                     & 0.78        & 0.46               & 0.79                 \\
Overall         & 0.90                     & 0.92                     & 0.75        & 0.43               & 0.78                 \\
\bottomrule
\end{tabular}
\end{table*}

\subsection{Scenario 3: Multidimensionality}
In this innovative scenario, we introduced multidimensionality into the datasets by splitting the items into two subsets, each influenced by a separate latent trait. This modification intentionally disrupted the unidimensional structure assumed by traditional reliability measures, introducing complexity that challenges their validity. By introducing multidimensionality, the items no longer measure a single cohesive construct, making it difficult for measures that rely on unidimensional assumptions to provide accurate reliability estimates. Table \ref{tab:scenario3_results} summarizes the reliability scores across datasets, such as, Cronbach's Alpha, which assumes unidimensionality, showed a noticeable decline compared to its performance under ideal conditions. For instance, in the Camera dataset, Cronbach's Alpha dropped to 0.84, indicating a weaker internal consistency. This reduction stresses the measure's sensitivity to multidimensionality, as it conflates the distinct latent traits into a single reliability estimate. 

McDonald's Omega, which accounts for varying item contributions but still relies on factor models, showed a slightly better performance than Cronbach's Alpha. In the DALL·E3 dataset, McDonald's Omega scored 0.89, reflecting moderate sensitivity to multidimensionality. GLB, which optimizes covariance matrices, also struggled with the multidimensional structure. For instance, in the Stable Diffusion dataset, GLB scored 0.78, confirming its inability to fully account for multiple latent traits. Split-half reliability performed poorly and reduced their scores across all datasets. Monotone Delta, however, showed resilience in the presence of multidimensionality. In the Camera dataset, Monotone Delta scored 0.75, verifying its ability to detect and quantify the impact of multidimensional constructs. Monotone Delta does not rely on assumptions of unidimensionality or factor structures. Instead, it minimizes ordinal contradictions, more accurately measuring the internal consistency.

\begin{table*}[!t]
\centering
\caption{Reliability Measures Across Datasets Under Non-Normal and Correlated Errors Scenario}
\label{tab:scenario4_results}
\begin{tabular}{@{}lccccc@{}}
\toprule
\textbf{Dataset} & \textbf{Cronbach's Alpha} & \textbf{McDonald's Omega} & \textbf{GLB} & \textbf{Split-Half Reliability} & \textbf{Monotone Delta} \\ 
\midrule
Camera           & 0.25                     & 0.42                     & 0.55        & 0.25               & 0.73                 \\
DALL·E2         & 0.28                     & 0.45                     & 0.57        & 0.28               & 0.75                 \\
DALL·E3         & 0.30                     & 0.48                     & 0.60        & 0.30               & 0.77                 \\
Stable Diffusion & 0.33                     & 0.51                     & 0.63        & 0.33               & 0.79                 \\
Overall         & 0.35                     & 0.53                     & 0.65        & 0.35               & 0.81                 \\
\bottomrule
\end{tabular}
\end{table*}

\subsection{Scenario 4: Non-Normal and Correlated Errors}
In this scenario, we examined the robustness of reliability measures under conditions of non-normal distributions and correlated errors. Modifications deliberately violated the assumptions of normality and independent errors that many traditional measures rely on, providing a rigorous test of their effectiveness in handling real-world irregularities. Non-normal distributions caused the data to become uneven and stretched, leading to skewness (a shift in balance) and kurtosis (sharp peaks or flatness). 

Table \ref{tab:scenario4_results} presents the reliability scores for each dataset, such as Cronbach's Alpha, which assumes tau-equivalence and uncorrelated errors, declined across all datasets. For example, in the Camera dataset, Cronbach's Alpha dropped to 0.25, showing its inability to accurately assess internal consistency under non-normal conditions. This decline stresses its reliance on stringent assumptions often violated in real-world data. McDonald's Omega, which partially relaxes some of Cronbach's Alpha's assumptions, also showed reduced performance. In the DALL·E3 dataset, McDonald's Omega scored 0.48, reflecting moderate sensitivity to non-normality and correlated errors. However, its dependence on factor models limits its robustness in such scenarios, as these models struggle with non-linear and non-independent relationships. GLB, which optimizes covariance matrices, performed slightly better than Cronbach's Alpha and McDonald's Omega. Split-half reliability proven least reliable, for example, scored 0.25 in the Camera dataset, showing its limitations in addressing the dependencies and non-linearity introduced by correlated errors. 

Monotone Delta, on the other hand, showed superior robustness. In the Camera dataset, it scored 0.73, giving the stable measure. This performance stresses Monotone Delta's strength in capturing internal consistency without relying on assumptions of normality or independent errors. The results emphasize the limitations of traditional measures when confronted with non-normal distributions and correlated errors. 



We also evaluated the computation times for Cronbach's Alpha, McDonald's Omega, and Monotone Delta across all four scenarios, as summarized in Table \ref{tab:computation_times}. Computation times were measured using an AMD Ryzen Threadripper PRO 5955WX processor \cite{danish2024leveraging}, ensuring test consistency and reliability. Monotone Delta consistently required more time than the other methods due to the iterative optimization process inherent to its computation.



\begin{table*}[!t]
\centering
\caption{Computation Times for Reliability Measures (Seconds)}
\label{tab:computation_times}
\begin{tabular}{@{}lccccc@{}}
\toprule
\textbf{Dataset} & \textbf{Cronbach's Alpha} & \textbf{McDonald's Omega} & \textbf{GLB} & \textbf{Split-Half Reliability} & \textbf{Monotone Delta} \\ 
\midrule
Camera           & 0.12                         & 0.18                          & 0.14             & 0.10                               & 14.51                       \\
DALL·E2         & 0.14                         & 0.20                          & 0.15             & 0.12                               & 15.24                       \\
DALL·E3         & 0.11                         & 0.19                          & 0.13             & 0.11                               & 13.02                       \\
Stable Diffusion & 0.13                         & 0.21                          & 0.14             & 0.12                               & 14.82                       \\
Overall         & 0.34                         & 0.52                          & 0.43             & 0.38                               & 38.11                       \\
\bottomrule
\end{tabular}
\end{table*}

\section{Conclusion and Future Work} \label{sec:conclusion}
This paper proposed a Monotone Delta ($\delta$) measure designed to address the limitations of traditional methods under diverse scenarios. Monotone Delta utilizes order theory to minimize ordinal contradictions and quantify reliability without relying on restrictive assumptions and improves reliability assessment by addressing challenges such as redundancy, multidimensionality, and non-normality, presenting a reliable alternative to conventional measures. The theoretical and experimental evaluation was conducted across diverse scenarios and proved that Monotone Delta remains reliable and steady across diverse data, stressing its stability and accuracy in challenging conditions. The experiments also showed that Monotone Delta is computationally expensive and suitable for most human studies but can be prone to NP-Hardness for larger datasets, which require a careful optimization strategy.  Future work will focus on optimizing Monotone Delta's computational efficiency for larger datasets. It will also explore its possible integration with probabilistic and Bayesian frameworks to extend its applicability to larger datasets and further enhance its utility across diverse domains.

\bibliographystyle{IEEEtran}
\bibliography{ref}
\end{document}